\pdfoutput=1
\documentclass[10pt,a4paper]{article}
\RequirePackage{latexsym,amsmath,amssymb}
\RequirePackage[dvipsnames,usenames]{color}

\usepackage{fullpage}

\usepackage[english]{babel}
\usepackage[latin1]{inputenc}
\usepackage[T1]{fontenc}
\usepackage[final]{showkeys} 
\usepackage[hidelinks]{hyperref}
\usepackage{cite}
\hypersetup{pdfauthor={M. Cadoni, M. Tuveri...},
            pdftitle={}
            }
\usepackage[]{cleveref}
\Crefname{equation}{Eq.}{Eqs.}
\Crefname{figure}{Fig.}{Figs.}
\crefformat{plural}{#2eqs.~(#1)#3}
\crefname{section}{Sect.}{Sects.}

\usepackage{amsfonts}
\def\lb{\label}
\def\be#1\ee{\begin{align}#1\end{align}}

\renewcommand{\le}{\leqslant}
\renewcommand{\leq}{\leqslant}

\def\lb{\label}

\title{\bf Long-Range Quantum Gravity}
\author{M.~Cadoni${}^{ab}$\thanks{E-mail: mariano.cadoni@ca.infn.it},
M. ~Tuveri${}^{b}$\thanks{E-mail: matteo.tuveri@ca.infn.it} \
and
A. P. ~Sanna${}$\thanks{E-mail: asanna9564@yahoo.it}
\\
\\
${}^a$\emph{Dipartimento di Fisica, Universit\`a di Cagliari}
\\
{\em Cittadella Universitaria, 09042 Monserrato, Italy}
\\
\\
${}^b$\emph{I.N.F.N, Sezione di Cagliari}
\\
{\em  Cittadella Universitaria, 09042 Monserrato, Italy}
\\
\\}
\begin{document}
\maketitle
\begin{abstract}

It is a tantalising possibility that quantum gravity (QG) states remaining 
coherent at astrophysical, galactic and cosmological scales could exist and that 
they could play a crucial role in understanding macroscopic gravitational effects.
We explore, using only general principles of General Relativity, quantum and 
statistical mechanics, the possibility of using long-range QG states to describe 
black holes. In particular, we discuss in a critical way the interplay between 
various aspects of long-range quantum gravity, such as the holographic bound, 
classical and quantum criticality and the recently proposed quantum thermal  
generalisation of Einstein's equivalence principle.  
We also show how black hole thermodynamics can be easily explained 
in this framework.

\end{abstract}

\section{Introduction}
Conventional wisdom asserts that quantum gravity effects may be relevant only at scales 
of the order of the Planck length $l_p  =\sqrt{\hbar G/ c^3}\sim 10^{-33} \ \text{cm}$. This simple result comes upon equating 
the Compton length associated with a self-gravitating object with its gravitational size, 
its Schwarzschild radius, $R_s$. On one hand, this would imply that only Planck-size black 
holes have an intrinsic quantum gravity nature. Conversely, astrophysical black holes, originated 
from the collapse of stars which, in turn, have proved their existence through gravitational waves 
detection~\cite{Abbott:2016blz} and from the first photo of their light ring~\cite{Akiyama:2019cqa}, 
can be described by means of classical gravity only. 
On the other hand, starting from the pioneering works of Hawking and Bekenstein~\cite{Bekenstein:1973,Hawking:1974,Hawking:1974sw}, 
we know that black holes are thermal objects, whose thermodynamic behavior cannot 
be explained by classical gravity alone. From a microscopic point of view, black holes entropy 
and temperature are a manifestation of the existence of some (yet unknown) internal degrees 
of freedom (DOF), i.e.~quanta, whose dynamics should be responsible for both 
the quantum and the thermodynamical  properties of these objects~\cite{Sakharov:1967pk,Jacobson:1995ab,Padmanabhan:2009vy,Jacobson:2015hqa,Padmanabhan:2016eld}. 
 
It is logically possible that the macroscopic behaviour of black holes is the result of a 
microscopic - Planck scale - QG theory, in the same way as black body emission  
and specific heat of solids are  manifestation of microscopic quantum mechanics. 
However, there are strong indications that this may not be the case. 
The area-law, which encodes the so-called \emph{holographic principle}, the 
information paradox~\cite{Hawking:2015qqa,Mathur:2009hf} and the fact that black 
holes have no hairs~\cite{Cardoso:2016ryw}, altogether they imply that the quantum 
characterization of a black hole must be done at the \emph{horizon scale}, $R_s$. 
Indeed, the holographic principle tells us that the would-be DOF making up black hole 
entropy are localized on the horizon, the information paradox concerns only near-horizon 
physics and black hole hairs can be fully expressed in terms of the size of the black hole.

The logically simplest solution to this puzzle is to assume that the black hole is made up of 
a large number of QG states, which remain coherent at the horizon scale. This quantum portrait, 
which describes black holes similarly to Bose-Einstein condensates (BEC), has been first proposed  
in Ref.~\cite{Dvali:2011aa} and has been extended to describe several macroscopic effects 
of gravity (the de Sitter (dS) universe, inflation, emergence of a dark force in galactic dynamics)
as long-range QG effects
~\cite{Mueck:2013mha,Dvali:2013eja,Das:2014agf,Oriti:2016acw,Casadio:2016zpl,Linnemann:2017hdo,Das:2018udn,Das:2018lwl,Tuveri:2019zor,De:2019fva,
Compere:2019ssx,Cadoni:2017evg,Casadio:2017cdv,Cadoni:2018dnd,Giusti:2019wdx,Tuveri:2019zor,Cadoni:2020izk}.
In this description, black holes are considered as critical systems, saturating 
a \emph{maximal packing condition}. 

However, this simple solution produces a certain tension with the equivalence principle 
of General Relativity (GR), which sees the black hole horizon as a place with nothing special. 
This tension can be solved by the formulation of the quantum \emph{generalized thermal equivalence principle (GTEP)}~\cite{Tuveri:2019zor}, 
which generalizes the classical equivalence principle of GR by postulating  
a fundamental relation between the temperature and the acceleration (the surface gravity).

In this essay we explore, using only general principles of GR, quantum and 
statistical mechanics, the possibility of using long-range QG states to describe black holes. 
In particular, we discuss in a critical way the interplay between various aspects of this description, namely the holographic principle, criticality and 
the peculiarities of the classical limit. 
We will show how the GTEP allows us to fit them in a nice and consistent description.  
Using a simple toy model to describe some aspects of black holes physics, 
we also show that the GTEP is a fundamental and unifying principle in QG. 
Indeed, we show that black hole thermodynamics can be fully derived from the combination of GTEP
with another principle chosen from $1)$ the holographic bound,
$2)$ classical and $3)$ quantum criticality. What is really interesting is that,
whereas the GTEP appears to be a fundamental principle, the others give three complementary 
description of a black hole, respectively as $1)$ a state  of maximal information, 
$2)$ a classical critical star and $3)$ a quantum critical state. 
We will be mainly concerned with black holes, but most of our considerations can be extended  
also to the case  of  the dS universe.

The structure of the paper is as follows. In Sect.~\ref{LRQG} we discuss the main features 
of our QG description, namely criticality, the holographic bound, the classical limit and the  
GTEP. In Sect.~\ref{sec:toy} we use a simple toy model for a black hole   
to see how they are intertwined among themselves and we show how the GTEP can 
be used as unifying principle to describe black holes. Finally, in Sect.~\ref{sect:conc} 
we present our conclusions.


\section{Long-range properties of quantum gravity }\label{LRQG}

Quantum mechanics can be responsible for the macroscopic behavior 
of several physical systems, like Bose-Einstein condensates, superfluids, 
superconductors and neutron stars. These are all quantum objects that 
can maintain quantum coherence even at macroscopic scales, so that both their  
macroscopic phenomenology and microscopic properties are a clear 
manifestation of quantum mechanics.  

There is growing evidence that a similar description could hold true also for 
gravitational systems like black holes and the dS universe~\cite{Witten:2001kn,Binetruy:2012kx,Mueck:2013mha,Das:2014agf,Oriti:2016acw,Casadio:2016zpl,Das:2018udn,Das:2018lwl,Tuveri:2019zor,De:2019fva}. 
Thus, long-range QG effects could be relevant for explaining 
macroscopic gravitational effects, such as, for instance, a non-Newtonian component 
of the acceleration in galactic and cosmological dynamics~\cite{Cadoni:2017evg, Casadio:2017cdv,Cadoni:2018dnd,Giusti:2019wdx,Tuveri:2019zor,Cadoni:2020izk}. 
The existence of long-range QG states allows one to circumvent the usual 
argument that QG effects are only relevant at scales of the order of the Planck 
length $l_p$. 
As in BECs, where the scale of quantum effects is the size of the condensate itself, 
in principle we can think that, for QG effects, it coincides with the Schwarzschild radius 
$R_s$ and the cosmological horizon $L$ for black holes and the dS universe, respectively~\cite{Verlinde:2016toy,Tuveri:2019zor}.
Moreover, in this picture, a \emph {mesoscopic} length-scales for QG may be generated 
at galactic level, if an interaction between dark energy and baryonic matter is assumed~\cite{Verlinde:2016toy,Cadoni:2018dnd}.
\medskip

The existence of a long-range QG regime is thus related to the peculiar features 
of gravity and in particular to: a) the peculiarity of the \emph{classical limit} in QG; 
b) the \emph {criticality} of the quantum gravitational systems; 
c) the \emph{holographic} character of gravity.
All these features are ruled by a single length scale, representing the size of the system.
In the following subsection, we will elucidate the interplay and the deep connection 
between these features, also arguing that a consistent description of them requires 
a generalization of Einstein's equivalence principle.

The simplest QG model for a black hole of mass $M$ is that of a 
coherent state of $N$ quanta of energy $\varepsilon$, with:
\be\lb{quanta}
\varepsilon=\frac{\hbar c}{R_s},\quad M{ c^2}= N\varepsilon.
\ee
In other words, a black hole is modeled as a cavity of length $\ell \sim R_S$, whose 
quantum degrees of freedom are quantum states in this cavity (see also~\cite{Casadio:2020ueb,Casadio:2020rsj} 
for a quantum description of black holes and gravity in terms of coherent states). 

Similar expressions hold true also for the dS universe, with $R_s$ 
replaced by $L$ and $M$ by the total (dark) energy inside the 
cosmological horizon. In this picture, gravity emerges to sustain the 
coherence of a quantum critical system at large scales 
(see also~\cite{Bruschi:2020xbm} for a discussion on this topic). 

%
\subsection{Criticality, the holographic bound and the classical limit}
For a generic gravitational system with radius $R$ and mass $M$, 
whose quantum description is given by ~\eqref{quanta}, the \emph{classical bound} 
\be\lb{star}
\frac{2GM}{R{c^2}}\le 1,
 \ee
translates into a maximally packing condition~\cite{Dvali:2010bf,Dvali:2010jz,Dvali:2011th}
\be \lb{mpc}
N\le \frac{{c^4} m_p^2}{2 \varepsilon^2},
\ee
where $m_p$ is the Planck mass. This quantum bound tells us that black 
holes are critical objects (see Sect.~\ref{sec:BHcritical}), although its physical 
origin is not completely clear. It does not seem to be a  consequence of the 
Heisenberg uncertainty principle alone, which only determines the relation between 
the size of the region and the energy of the quanta, $\varepsilon=\hbar c/R_s$ 
or, equivalently, $M_{BH}= m_p^2 c^2/(2 \varepsilon)$. 
Thus, Eq.~\eqref{mpc} seems to have a genuine long-range QG origin in the 
form of some effective repulsive interaction which, in turns, stabilizes the black hole.
\medskip 

The criticality bound~\eqref{mpc} can be written, upon using Eq.~\eqref{quanta}, as
\be \lb{hb}
N\le \frac{R_s^2}{l_p^2},
\ee
which is the well-known \emph{holographic bound}~\cite{Susskind:1994vu}, limiting  
the quantity of information we can store inside a sphere with area $4\pi R_s^2$. 

Although the two bounds~\eqref{hb} and~\eqref{mpc} are equivalent, they are 
conceptually independent and have a completely different origin and interpretation. 
Whereas the former has an informational nature - its origin being related to the 
Bekenstein-Hawking formula for black hole entropy- the latter has a dynamical 
nature. 
Furthermore, Eq.~\eqref{hb} is intrinsically holographic, whereas Eq.~\eqref{mpc} 
does not give any hint about the localization of the $N$ quanta - they may 
be as well localized inside the sphere or on its boundary.
We will further discuss this point in Sect.~\ref{sec:toy}.
\medskip 

Even if they can be described as quantum critical systems, black holes 
(and the dS universe) are solutions of Einstein's GR and thus have also a 
classical interpretation, independent from their microscopic structure. However, the 
transition from the quantum to the classical description is highly non-trivial. 
It involves both the usual classical limit~$\hslash \to 0$ and a {\it classicalization} 
process related to the number of quantum coherent states of the system, $N$, 
and it occurs when $N\to \infty$~\cite{Dvali:2011aa,Dvali:2012en,Dvali:2012rt}, 
as in BECs. Macroscopic quantum phenomena are thus associated either to a 
system with a large number of states or to a quantum state occupied by a large 
number of particles. This is surely quite intuitive for a black hole, where quantum 
states have macroscopic size of the order of its Schwarzschild radius. 
From a technical point of view, this is a typical feature of systems where conformal 
invariance arises~\cite{Cadoni:2006ww}.

A key point is that, in a statistical mechanical description, any quantum system is 
characterized by thermal macroscopic quantities reflecting its microscopic 
structure, like its temperature and entropy. 
Thus, the bridge between the classical and quantum description is provided by 
black hole thermodynamics, which has been widely used to give a coarse grained 
thermal description of black holes.

\subsection{Generalized Thermal Equivalence Principle (GTEP)}
The thermodynamic behavior of the Schwarzschild black hole is fully 
characterized by its temperature $T$, mass $M$ and entropy $S$:
\be\lb{tp}  
T=\frac{\hbar c}{4 \pi R_s},\quad M=\frac{R_s { c^2}}{2G},\quad S=\frac{\pi}{l_p^2} R_s^2,
\ee
from which the first principle of thermodynamics $dM=TdS$ follows immediately 
by considering a classical process that changes the black hole radius $R_s$.  
Thus, black hole thermodynamics is fully defined in terms of fundamental constants 
($\hbar,c,G$) and one single macroscopic quantity: the black hole radius $R_s$. 
This is the thermodynamic manifestation of black holes' criticality, or, equivalently, 
of the maximally packing condition. 
More generally, Eqs.~(\ref{tp}) imply that 
black hole thermodynamics is a manifestation of quantum physics acting 
at horizon scale. 

Furthermore, we see that $\hbar$ does not enter in the mass/radius relation. 
This latter can be thought as the analogous to compactness relation for a star 
and, in principle, can have a classical explanation. 
On the other hand, the entropy/radius relation depends on both $\hbar$ and $G$ 
and can therefore have only a QG origin. 
In terms of the coherent states building the black hole, it represents the maximal 
amount of information which can be codified in a spherical region of radius $R_s$ 
via the holographic bound~\eqref{hb}.
Finally, the relation between the temperature and $R_s$ is rather mysterious when 
considered in the classical limit. In terms of quantum mechanics, it reflects the fact 
that the temperature measures the average energy of the quanta, i.e. $T\sim \varepsilon=\frac{\hbar c}{R_s}$. 
This is particularly evident when one rewrites it in terms of the surface gravity $a$,
\be\lb{tsg}   
T= \frac{\hbar}{2\pi c} a,  
\ee
where:
\be\lb{tsg1}   
a= \frac{c^2}{2R_s}.  
\ee
This temperature/acceleration relation is quite puzzling, particularly considering that 
it does not depend directly of the strength of the gravitational interaction $G$. Another 
puzzling point is the tension that exists between the equivalence principle of GR, which 
implies that the black hole horizon is not a special place in spacetime,
and the long-range QG description aiming to explain the black hole in terms of quantum 
states with horizon size.

The simplest solution of these puzzles, which also allows to reconcile the classical and 
quantum perspectives in describing black hole physics, is to explain Eq.~\eqref{tsg} 
as a consequence of a \emph{Generalized Thermal Equivalence Principle} (GTEP)~\cite{Tuveri:2019zor}.
It has been formulated as the generalization of Smolin's quantum version of 
the universality of free fall (the thermal equivalence principle)~\cite{Smolin:2017kkb}, 
which is based on the Deser-Levin formula~\cite{Narnhofer:1996zk,Deser:1997ri,Jacobson:1997ux}. 
The GTEP asserts that whenever we have a thermal ensemble at temperature $T$ of 
quantum gravity degrees of freedom, the macroscopic acceleration produced on a test 
mass is given by the Deser-Levin formula
\be\lb{GTEP}
a=\frac{2\pi c}{\hslash}\sqrt{T^2-T^2_{dS}},
\ee 
where $T_{dS}$ is the dS temperature. 
One can easily check that Eq.~\eqref{tsg} simply follows from Eq.~\eqref{GTEP} when 
$T\gg T_{dS}$. The GTEP has been used to explain galactic dynamics and in particular 
the Tully-Fisher relation, without assuming the existence of dark matter~\cite{Tuveri:2019zor}.
The connection with the QG description of black holes can be obtained using Eq.~\eqref{quanta}
into Eq.~\eqref{tsg1} to find a nice relation between the energy of the quanta and the acceleration:
\be\lb{dl2}
a=\frac{c}{2\hbar}\varepsilon. 
\ee
The same result can be obtained in the corpuscular model of black holes of 
Refs.~\cite{Dvali:2013eja,Casadio:2016zpl,Cadoni:2017evg,Cadoni:2018dnd,Dvali:2012rt}. 

Conversely, Eqs.~\eqref{dl2} and~\eqref{quanta} can be used to derive Eq.~\eqref{tsg1}, i.e. 
to understand the peculiarity of the classical limit involved in describing black holes as classical 
objects. Indeed, we see that the surface gravity~\eqref{tsg1} has a quantum origin, but 
its  do not scale away in the limit $\hslash\to 0$  as it should be the  case in  standard 
quantum mechanics. This is because $\hslash$  appears in the numerator of $\varepsilon$ (see Eq. ~\eqref{quanta}) but in the denominator of   Eq. ~\eqref{dl2}.

Notice that Eq.~\eqref{tsg1} is the result of {\em both} the GTEP and criticality. 
To see this, let us consider a \emph{non-critical} classical gravitational system, 
i.e. a system for which the bound~\eqref{star} is not saturated.
We consider a spherically symmetric configuration of energy $E=Mc^2$, 
inside a sphere of radius $R$, which may or not coincide with the physical radius 
of the system. We also assume that the gravitational system allows for a microscopic 
long-range QG description in terms of $N$ degrees of freedom characterized by the 
temperature $T$ (derived upon a suitable coarse grained operation). 
Notice that, in this situation, Eqs.~\eqref{quanta} do not necessarily hold.

Furthermore, we assume: a) the validity of GTEP, given by Eq.~\eqref{GTEP} 
with $T\gg T_{dS}$, i.e. $a=\frac{2\pi c}{\hslash}T$; 
b) the saturation of the holographic bound~\eqref{hb}, $N=R_s^2/l_p^2$; 
c) an equipartition rule for the energy inside the sphere: $ E=Mc^2=  \frac{1}{2} N T$.
One can easily check that Newton's law $a_N = GM/R^2$ simply follows from putting 
together a), b) and c). 
Only when the criticality condition~\eqref{star} is saturated, $a_N$ takes the form~\eqref{tsg1}.  
\medskip
  
The GTEP also applies to dS universe. 
In this case, Eq.~\eqref{dl2} gives the well-known relation between the dS temperature 
and the cosmological acceleration $H$~\cite{Narnhofer:1996zk,Deser:1997ri}: 
$a=H=\frac{2\pi c}{\hbar }T_{dS}=\frac{c^2}{ L }$.
The dS universe can be thought as an ensemble of quanta with typical energy
$\varepsilon= \hbar c/L$, so that we find $a=\frac{c}{\hbar}\varepsilon$,
which is similar to Eq.~\eqref{dl2} up to a factor of $2$.

\section{A black hole toy model}
\lb{sec:toy}
In the previous Section, we have seen how quantum and classical properties 
of black holes, including their thermodynamics, can be fully explained using 
three main principles: criticality, the holographic bound and the GTEP.
Two important, albeit related,  questions arise: are these principles logically 
independent one from the other or can one of them be derived from the others? 
Is one of them more fundamental than the others?

In this Section we will try to settle down these issues  by building a simple toy 
model to describe black holes as quantum gravitational systems, with a large 
number of internal degrees of freedom, $N\gg 1$, with typical energy and 
mass given by Eq.~\eqref{quanta}.

\subsection{Black hole as a critical star}
Let us assume, beyond the validity of GTEP, that a black hole has a classical 
description in term of GR. We define the Schwarzschild radius as $R_s=2GM/c^2$. 
This means that the classical criticality bound~\eqref{star} is saturated, namely the 
black hole is considered as a sort of critical star (see also~\cite{Casadio:2019tfz,Calmet:2019eof}). 
Putting together Eq.~\eqref{quanta} and the definition of the Schwarzschild radius, 
we find that the number of degrees of freedom in the black hole cavity scales as the 
area, i.e. it scales holographically in the sense previously described:
\begin{equation}\label{Ndof}
N\sim\frac{{ c^3}}{G\hslash}R_S^2.
\end{equation}
This allows us to find that also the entropy of the black hole scales holographically 
as its area:
\begin{equation}\label{Sbh}
S\sim N\sim \frac{R_S^2}{l_P^2}.
\end{equation}
By using the GTEP, we can easily find the black hole temperature and, more in general, 
characterize its thermodynamics. Since in the case under study we expect that $T\gg T_{dS}$, 
putting together Eq.~\eqref{GTEP} and Eq.~\eqref{tsg1}, we find the expected result:
\begin{equation}\label{Tbh}
T=\frac{\hslash c}{4\pi R_S}.
\end{equation}
Thus, it is possible to derive black holes thermodynamics by considering them 
as critical stars in GR and assuming the validity of the GTEP.
In this case, holography appears as a consequence of these twos rather than 
a first principle.

\subsection{Black hole as a critical quantum system}
\label{sec:BHcritical}
Let us now consider a black hole of mass $M$ as a quantum critical system.
In this case, we assume the validity of the GTEP and the saturation of the bound~\eqref{mpc}:
\begin{equation}\label{N_LRQ}
N=\frac{m_p^2 { c^4}}{2}\frac{1}{\varepsilon^2}.
\end{equation}
The total mass of the system is:
\begin{equation}\label{M_LRQ}
M=\frac{m_p^2 c^2}{2}\frac{1}{\varepsilon}.
\end{equation}
Eq.~\eqref{N_LRQ} states that for a given mass $M$, the number of quantum degrees 
of freedom cannot vary freely, rather it depends on the energy $\varepsilon$. 
In a system of finite size, we can pack many states of small energy or few states of 
large energy. This implies that, at Planck scales, black holes have $N\sim 1$, 
whereas astrophysical black holes have $N\gg 1$. This means that they satisfy 
the maximally packing condition described in Sect.~\ref{LRQG}. 
Thus, in a given region of size $R_S$ and corresponding quanta of energy 
$\varepsilon=\hslash{ c}/R_S$, we can put $N\leq m_P^2{c^4}/\varepsilon^2$ quanta. Once 
the bound $N=m_p^2{ c^4}/\varepsilon^2 $ is saturated we cannot put other quanta in that 
region without enlarging it. Thus, black holes are classically characterized 
by the condition $M=Rc^2/2G$, which is analogous to the condition $N\varepsilon^2=m_p^2 {c^4}/2$ 
in the quantum portrait~\cite{Dvali:2011aa,Dvali:2012en,Dvali:2012rt,Dvali:2013eja}. 
The saturation of the holographic bound~\eqref{hb} easily follows from Eq.~\eqref{N_LRQ} 
upon using Eq.~\eqref{quanta} and the Bekenstein-Hawking formula for the entropy 
in~\eqref{Sbh}. On the other hand, the black hole temperature~\eqref{Tbh} still follows 
from the GTEP, whereas the mass/radius relation is now obtained from~\eqref{M_LRQ}. 

In this derivation, black hole thermodynamics follows in a straightforward way from the 
GTEP and the quantum criticality condition (\ref{N_LRQ}). 

\subsection{Black holes as states of maximal information}
In the previous subsection we have seen that the holographic bound~\eqref{hb} is equivalent, 
upon the use of Eq.~\eqref{quanta}, to the criticality bound~\eqref{mpc}. It follows that 
we can derive black hole thermodynamics from the GTEP and the requirement of the saturation 
of the holographic bound~\eqref{hb}. 
The Bekenstein-Hawking entropy $S$ follows directly from the latter, the Hawking temperature 
follows from the former, whereas the expression of the mass $M$ follows from the second Eq.~\eqref{quanta}.  

We stress again that, despite the equivalence between the holographic bound~\eqref{hb} and 
the criticality bound~\eqref{mpc}, they have a completely different conceptual meaning. 
Whereas the latter has an informational nature, the former has a fully dynamical 
meaning.

It is interesting to see how, in the last two derivations of black hole thermodynamics, the black hole 
mass $M=R_sc^2/2G$ is obtained by summing up the energy of $N$ quanta (see Eq.~\eqref{quanta}). 
The mass $M$ is a classical observable and cannot vanish in the $\hbar\to 0$ limit. 
Indeed, $\hbar$ cancels out by putting together Eqs.~\eqref{M_LRQ} and~\eqref{quanta}. 
This is completely analogous to the $\hbar$ classical limit cancellation, which occurs for the surface 
gravity. We expect this $\hbar$ cancellation to hold not only for the mass but also for all black hole hairs.

\section{Conclusions}
\lb{sect:conc}

In this paper we have explored the possibility that quantum gravity states remaining 
coherent at astrophysical  scales could be used to describe black 
holes. They can be seen as quantum macroscopic critical objects, which are 
characterized by a typical length scale (their radius) which determines all their features 
in terms of $N$ coherent states, i.e. quanta with a typical energy and temperature 
determined by the black hole radius.
In the critical phase, it is not possible to put any further quanta in the system without 
changing its size. This is related to the saturation of the informational (entropy) bound 
which, in the case of black holes, appears as a holographic bound.   
One interesting point is that the same features are shared by the de Sitter universe. 

We have also discussed the interplay between the various aspects of the long-range 
quantum gravity description of black holes, namely the holographic principle, criticality 
and the peculiarities of the classical limit.
We have seen that a quantum, thermal extension of Einstein's equivalence principle, 
which we called GTEP, allows for a nice, consistent and unifying description of black 
holes. 
In particular, we have seen how it is possible to derive black hole thermodynamics 
starting from three main principles: classical or quantum criticality, holography and 
the GTEP. 
The GTEP seems to be more fundamental than the others. Indeed, the choice of one 
instead of the other remaining principles leads to different complementary description 
of a black hole, i.e. as a state of maximum information, as a classical or a quantum 
critical state, respectively.

Let us conclude with the main caveats of our 
description of black holes in terms of long-range QG.
The discussion we have presented in this paper has a rather speculative character.  
Although based on general features of General Relativity, quantum and statistical 
mechanics, it suffers from the fact that until now, we do not have a well-defined quantum 
theory of gravity.  
Only in that framework, notions like "QG long-range coherent states" we have used 
in this paper would have a well-defined meaning.
On the other hand, it is quite clear that any formulation of a quantum theory of gravity 
requires ingredients bridging between GR and quantum mechanics. The basic principles 
on which we have built our discussion, namely the holographic principle, criticality and 
the GTEP, have to be considered as bridging principles in this direction. 
One of the main results of our discussion is that the GTEP seems to work at a more 
fundamental level than the others.

Another question we have in mind (probably strongly related to the previous one) is 
about the physical mechanism underlying the stability of black holes. 
Classicalization and criticality should prevent the formation of singularities allowing 
the formation of stable critical (quantum gravitational) macroscopic systems made 
by a large number of quantum coherent states with energies of the order of $1/\ell$, 
where $\ell$ is the typical size of the system. However, if black holes can be 
described by something similar to a BEC, why does the condensate remain stable? 
Gravity is an attractive force, and in principle, a condensate of gravitons should collapse 
to form a singularity if some repulsive potential does not arise to compensate the gravitational 
interaction and stabilize the system. 
Standard BECs are stabilized by repulsive forces typical of Coulombian interactions between 
atoms in the critical phase. This is not the case of gravitational systems, where no repulsive 
components of the force arise in the system and the only fundamental force governing them 
seems to be  gravity. 

A different way to formulate the previous question 
is to ask about the physical origin of the holographic and the criticality bounds, 
in Eqs.~\eqref{hb} and~\eqref{mpc}, respectively. 
Have they a purely informational origin or have they a fundamental dynamical nature?
It is quite clear that the answer to these questions is one of the main challenges of any 
quantum theory of gravity. 



\providecommand{\href}[2]{#2}\begingroup\raggedright\endgroup

\end{document}